\documentclass[usenatbib]{emulateapj}

\shorttitle{The Assembly History of Galaxies}
\shortauthors{A. R. Hill et al.}
\begin{document}
\title{The Mass Growth and Stellar Ages of Galaxies: Observations versus Simulations}

\author{Allison R. Hill$^{1}$, Adam Muzzin$^{2}$, Marijn Franx$^{1}$, Danilo Marchesini$^{3}$}
\affil{$^{1}$Leiden Observatory, Leiden University, P.O. Box 9513, 2300 RA,
Leiden, The Netherlands}
\affil{$^{2}$Department of Physics and Astronomy, York University, 4700 Keele St., Toronto, Ontario, Canada, MJ3 1P3}
\affil{$^{3}$Physics and Astronomy Department, Tufts University, 574 Boston Avenue, Medford, MA, 02155, USA}

\email{hill@strw.leidenuniv.nl}

\begin{abstract}

Using observed stellar mass functions out to $z=5$, we measure the main progenitor stellar mass growth of descendant galaxies with masses of $\log{M_{*}/M_{\odot}}=11.5,11.0,10.5,10.0$ at $z\sim0.1$ using an evolving cumulative number density selection. From these mass growth histories, we are able to measure the time at which half the total stellar mass of the descendant galaxy was assembled, $t_{a}$, which, in order of decreasing mass corresponds to redshifts of $z_{a}=1.28, 0.92, 0.60$ and $0.51$. We compare this to the median light-weighted stellar age $t_{*}$ ($z_{*} = 2.08, 1.49, 0.82$ and $0.37$) of a sample of low redshift SDSS galaxies (from the literature) and find the timescales are consistent with more massive galaxies forming a higher fraction of their stars ex-situ compared to lower mass descendants. We find that both $t_{*}$ and $t_{a}$ strongly correlate with mass which is in contrast to what is found in the EAGLE hydrodynamical simulation which shows a flat relationship between $t_{a}$ and $M_{*}$. However, the semi-analytic model of \citet{henriques2015} is consistent with the observations in both $t_{a}$ and $t_{*}$ with $M_{*}$, showing the most recent semi-analytic models are better able to decouple the evolution of the baryons from the dark matter in lower-mass galaxies.

\end{abstract}


\section{Introduction}
\label{sec:intro}

Inferring the assembly history of present-day galaxies is challenging. It requires accurately linking progenitor to descendant, a process which is obfuscated by the fact that we only ever observe a galaxy at one snapshot in time. However, by using mass-complete censuses of galaxies at different redshifts and observing how populations of galaxies move through various parameter spaces (i.e., SFR, sSFR, central surface mass density, central stellar velocity dispersion, number density, etc.), one can begin to connect descendant galaxies to their likely progenitor population. 

By tracing galaxy evolution using a variety of the aforementioned parameters, observational studies are united in the finding that massive galaxies assemble most of their stellar mass before low mass galaxies, indicative of baryonic 'down-sizing' \citep[e.g.,][]{perezgonzalez2008, marchesini2009, behroozi2013, muzzin2013b, gonzalezdelgado2017}. This is consistent with analyses of the stellar populations of local galaxies, which find that more massive galaxies are host to older stellar populations \citep[e.g.,][]{kauffmann2003, gallazzi2005, thomas2010}

In contrast to observations, semi-analytic models (SAMs) and hydrodynamical simulations do not share the same consistency. Although both hydrodynamical simulations and semi-analytic models reproduce the positive correlation of stellar age with stellar mass, they differ in their predictions for when that mass assembled. Recent SAMs predict massive galaxies forming earlier than their low mass counterparts \citep[e.g.,][]{henriques2015}, in contrast to recent hydrodynamical simulations who show either a flat relationship between assembly time (the time at which 50\% of the mass was assembled) and stellar mass \citep{qu2017}, or a weak positive correlation \citep{sparre2015}. 

Although these models are inconsistent with each other on trends of stellar mass with assembly, they do all predict a higher fraction of the stars in massive galaxies were formed ex-situ \citep[e.g.,][]{rodriguezgomez2016, qu2017, mundy2017}. This picture is consistent with observations that indicate mergers are an important avenue of mass growth in massive galaxies since $z\sim1$ \citep[e.g,][]{newman2012, hill2017}. However, the role of mergers in the mass growth of lower mass galaxies remains uncertain. 

In this Letter, we endeavour to draw a direct observational comparison between the assembly time and the mass-weighted stellar age of galaxies, and demonstrate more concretely the relationship between galaxy stellar mass and the fraction of ex-situ stars. We also compare these timescales to the EAGLE simulation as well as the recent SAM of \citet{henriques2015} (hereafter H2015).

Unless otherwise specified, all ages and assembly times are for galaxies corresponding to a references redshift of $z=0.1$, with all ages reported in lookback times. We assume a $\mathrm{\Lambda}$-CDM cosmology ($H_{0}=\mathrm{70~kms^{-1}Mpc^{-1}}$, $\Omega_{M}=0.3$, and $\Omega_{\Lambda}=0.7$).

\begin{figure*}[ht]
  \begin{center}
  \includegraphics[width=\textwidth]{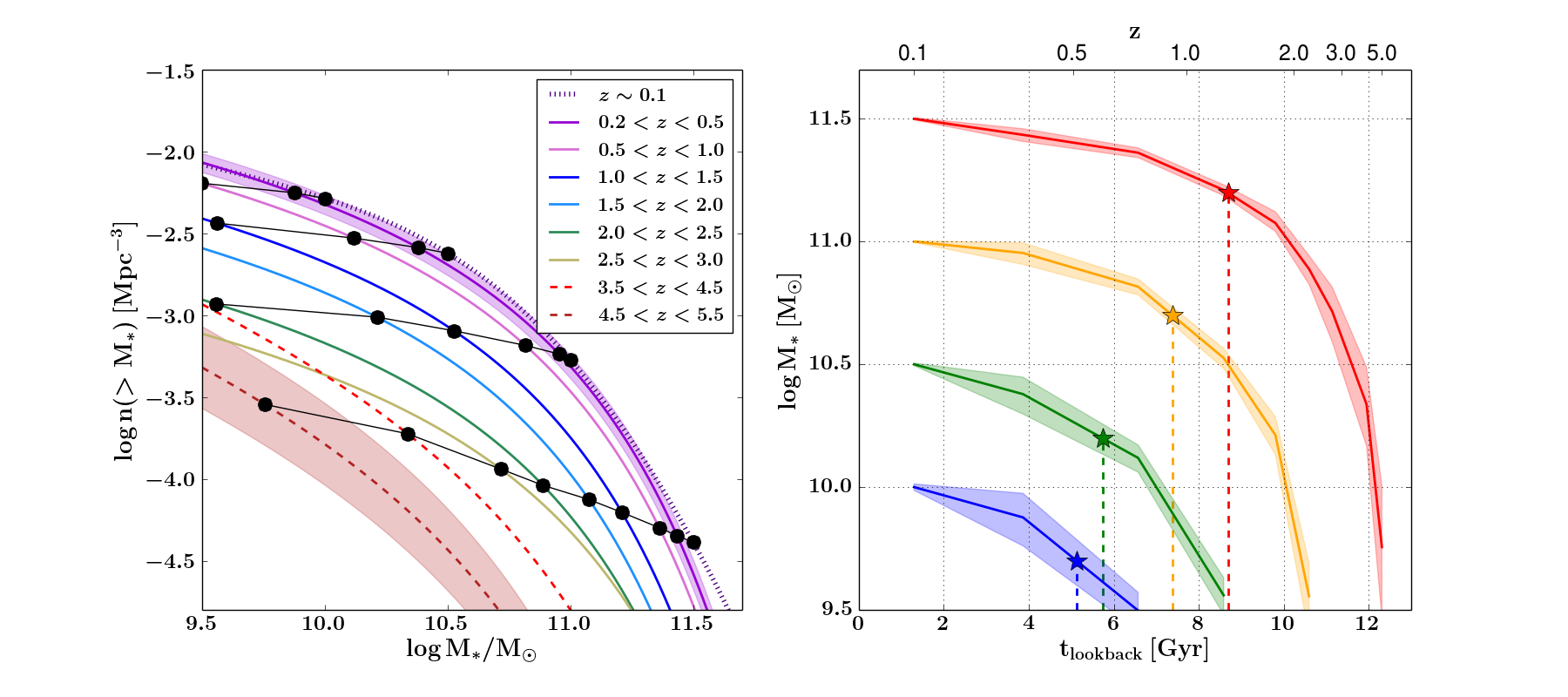}
  \caption{\textit{Left}: The cumulative number-density as a function of stellar mass at different $z$. Solid, dashed and dotted lines indicate the mass functions of \citet{muzzin2013b}, \citet{grazian2015}, and \citet{bernardi2017} respectively, with colour indicating the redshift. Uncertainties in the mass functions take into consideration the uncertainties in the photo-$z$'s, SFH and cosmic variance. For clarity, only the uncertainties for the highest- and lowest-$z$ are shown (as the uncertainties monotonically increase with $z$) . Black circles indicate the cumulative number density selection of \citet{behroozi2013} for four different descendent masses at $z\sim0.1$ ($\log{M_{*}/M_{\odot}}=11.5,11.0,10.5,10.0$). \textit{Right:} The corresponding mass evolution of the descendants considered in the left panel. Shaded regions indicate the uncertainty in the progenitor mass from the uncertainties in the mass functions. We trace the progenitors of four different descendant masses at $z\sim0.1$. Also plotted are the assembly times ($t_{assembly}$; coloured stars), which are the times at which half the final descendant mass is assembled.}
  \label{fig:progenitors}
  \end{center}
\end{figure*}

\section{Analysis}

\subsection{Measuring the assembly times}

To estimate the assembly time ($t_{a}$) for a galaxy, we must first determine a mass assembly history. The first challenge to analyzing the mass evolution of present day galaxies is properly identifying their progenitors. There are several methods to do this, e.g., by inferring the mass growth from the evolution of the SFR-mass relation \citep[e.g,][]{patel2013b}, selection via central surface-mass density \citep[e.g.][]{vandokkum2014}, selection via fixed central velocity dispersion \citep[e.g.,][]{bezanson2012}, and the evolution of the stellar mass function \citep[e.g.,][]{perezgonzalez2008, marchesini2009, muzzin2013b} (among others). The simplest and most appropriate method to derive the progenitor masses of galaxies is through cumulative number density selection. This method begins with the simple assumption that cumulative density would remain constant if there were no mergers or scatter in assembly; the evolution in the cumulative density due to these effects can be predicted robustly from models \citep{behroozi2013}. These predictions have been tested and verified  against more detailed simulations which accurately recover the median mass evolution\citep[e.g.,][]{torrey2015, clauwens2016, wellons2016b}. This method is the only method which can give a fair estimate from the evolution of the mass function alone, i.e., it does not need any detailed modelling to the full $f(M_{*}, SFR, merger~rate)$ distribution of galaxies.

In Figure~\ref{fig:progenitors} we show the number density, and progenitor mass evolution for four different descendant masses of  $\log{M_{*}/M_{\odot}}=11.5,11.0,10.5,10.0$ at $z\sim0.1$. As in \citet{hill2017}, we utilize the mass functions of \citet{muzzin2013b}, \citet{grazian2015} (with the addition of \citealt{bernardi2017} to extend to $z\sim0.1$) to translate the number densities from \citet{behroozi2013} into galaxy stellar masses as a function of redshift (left panel, Figure~\ref{fig:progenitors}). The regular evolution of the implied progenitor mass as a function of redshift in the right-panel highlights the quality of the input mass functions. Also indicated in the right panel of Figure~\ref{fig:progenitors} are the assembly times, $t_{a}$, the points at which half the final stellar masses were assembled. For our progenitors selection, this corresponds to an assembly redshifts (in order of decreasing stellar mass) of $z_{a} = 1.28, 0.92, 0.60, 0.51$. In this plot we see a clear trend towards baryonic cosmic `down-sizing', with the most massive galaxies assembling half their stellar mass earlier. 

\subsection{Measuring the stellar ages}
\label{sec:age_stellar}

To compare $t_{a}$ to the present-day age of the stellar populations in those galaxies, $t_{*}$, we take the light-weighted ages from \citet{gallazzi2005} ($t_{*,LW}$) which were measured from a subsample of 44 254 SDSS galaxy spectra. This subsample was chosen such that the median S/N per pixel was greater than 20, in order to accurately, and simultaneously model both the age and metal sensitive spectral indices such as $H\beta, H\delta_{A}, H\gamma_{A}$, $D4000$, and $[Mg_{2}Fe]$. They also were careful to exclude galaxies at redshifts which deviated substantially from the Hubble flow, resulting in a redshift range of $0.005<z<0.22$, with a median redshift of $z=0.13$. Extensive and careful modelling, using a library of 150 000 Monte Carlo realizations which cover a wide parameter space of plausible star formation histories, were used to accurately determine both age and metallicity as well as quantify the magnitude of the errors on these derived quantities. A full description of their methods can be found in \citet{gallazzi2005}.

For a galaxy with a given $M_{*}$, we take the median $t_{*,LW}$ (see Table~2 in \citealt{gallazzi2005}). As $t_{*,LW}$ is a median value, the formal errors are small (fractions of a percentage point), so we do not include those errors. However, as the SDSS fibres impose an aperture, there is potential for errors resulting from age gradients, especially in the larger galaxies. A recent analysis of age gradients in SDSS galaxies by \citet{goddard2017} found gradients at a level of $\sim0.1~dex/R_{e}$ from the centre to $1.5R_{e}$. This translates to an aperture correction of approximately $10\%$, which we use as a conservative error estimate in the median $t_{*,LW}$.

Light-weighted ages are biased towards younger stellar populations, as young stars dominate the optical emission where many age sensitive indices are measured \citep[see][]{kauffmann2003}. A more representative $t_{*}$ metric is the mass-weighted age, $t_{*,MW}$. Since the SFH is not known for these galaxies, we generate stellar-mass dependent corrections to $t_{*,MW}$ using the differences between the mass-weighted ages and r-band weighted ages from both H2015 (available in their catalog) and EAGLE (James Trayford, private communication) and apply it in the following way:

\begin{equation}
t_{*,MW,G05} = t_{*,LW,G05} + (t_{*,MW,sims} - t_{*, LW,sims})
\label{eq:t_gall}
\end{equation}

\begin{figure}
  \begin{center}
  \includegraphics[width=\linewidth]{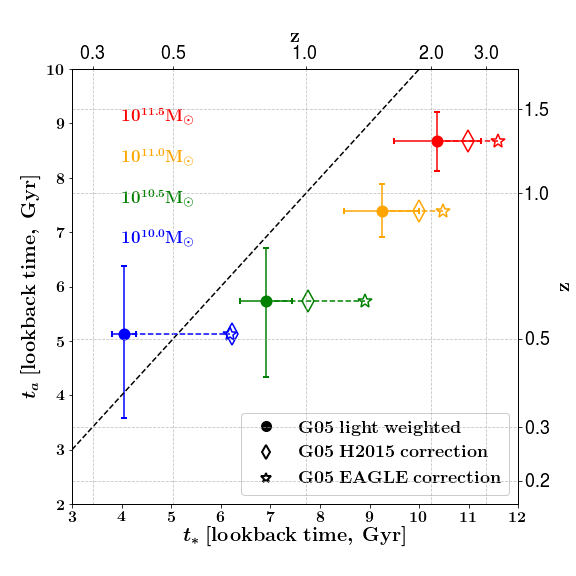}
  \caption{The assembly age, $t_{a}$, plotted as a function of the stellar age, $t_{*}$. $t_{a}$ is defined as the age at which half the stellar mass was assembled, as determined from the stellar-mass evolution tracks in the right panel of Figure~\ref{fig:progenitors}, with the errors estimated from uncertainties from the stellar mass functions.  The stellar ages are the median light-weighted ages (filled-circles) taken from \citet{gallazzi2005}, with the errors representing the expected uncertainty resulting from age gradients (see text for details). We have also estimated a mass-weighted age correction to the light-weighted ages using corrections measured from both EAGLE and H2015 (details can be found in the main text). We see a positive correlation between $t_{*}$, $t_{a}$ and mass, with the most massive galaxies assembling first.}
  \label{fig:ages}
  \end{center}
\end{figure}

Figure~\ref{fig:ages} shows $t_{a}$, $t_{*,MW,G05}$ for both H2015 and EAGLE, and $t_{*LW,G05}$ for all of our descendant galaxy masses. We see a range of assembly times, from $\sim5~\mathrm{Gyr}$ at the low-mass end, to almost $9~\mathrm{Gyr}$ for our highest mass bin. The span is larger in $t_{*}$ where we see a range of $\sim5-11~\mathrm{Gyr}$. We see all values are consistent with $t_{a}<t_{*,MW}$ which confirms our results are physical. We observe $t_{*}-t_{a}$ increasing with stellar mass, which suggests a higher fraction of the stars in massive galaxies are formed ex-situ than at lower masses. When comparing $t_{*}-t_{a}$ to the ex-situ fractions of the H2015 SAMs, they imply an ex-situ fraction of between $3-33\%$ for $\log{M_{*}}\geq10.5$ and between $1-33\%$ for $\log{M_{*}}=10.0$. This finding is consistent with other observational studies \citep[e.g, most recently,][who use sub-halo abundance matching to find a $M_{*}\sim5\times10^{11}\mathrm{M_{\odot}}$ galaxy has $\sim36\%$ of their mass formed ex-situ compared to only $\sim2.4\%$ for Milky-Way mass galaxies]{rodriguezpuebla2017}. This trend is also seen simulations (see Sec.~\ref{sec:intro} and references therein).

\subsection{Comparison to Simulations}
\label{sec:simualtions}

\begin{figure*}[ht]
  \begin{center}
  \includegraphics[width=\linewidth]{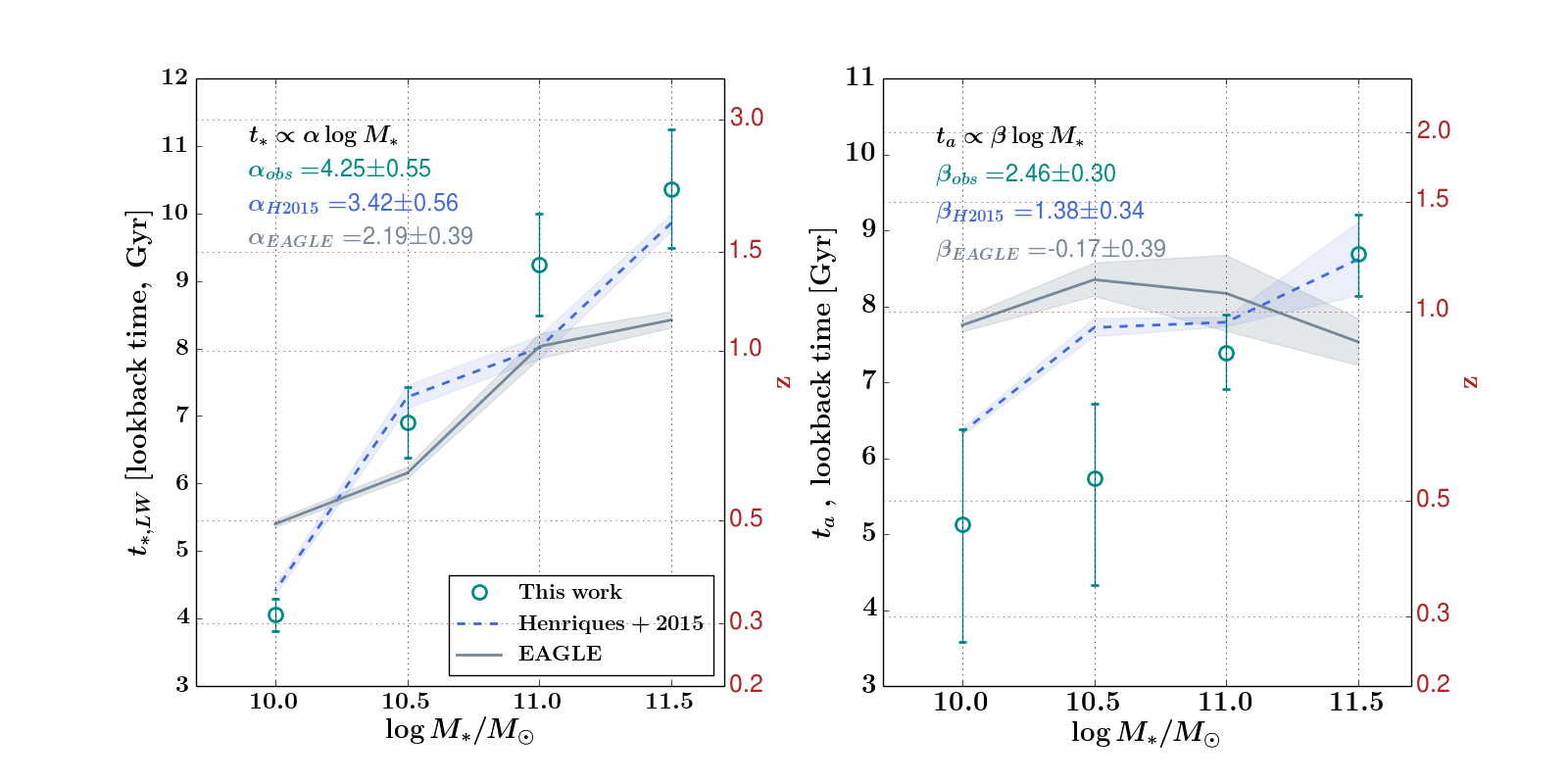}
  \caption{The stellar age ($t_{*}$; left panel) and assembly age ($t_{a}$; right panel) plotted as function of their final stellar masses. The turquoise circles are the values determined from observations (as in Figure~\ref{fig:ages}), solid grey lines are median values from the EAGLE simulation Ref-L100N1504, and dashed blue lines are from H2015. with the shaded regions around these lines representing the error implied from bootstrapping the samples.  We see the same positive correlation between $t_{a}$ and $t_{*}$ with stellar mass as implied in Figure~\ref{fig:ages}. The simulations also show this positive correlation with $t_{*}$ and $M_{*}$, with H2015 better matching (to within $2\sigma$) the steep dependence than EAGLE, which has a flatter relationship than the observations imply. In the right panel, we see the same flat relationship between $M_{*}$ and $t_{a}$ in EAGLE as found in \citet{qu2017} which does not match the observations, and is in fact consistent with a slope of 0. The $t_{a}$ measured from H2015 are in good agreement with the observations (to within $2\sigma$).}
  \label{fig:ages_masses}
  \end{center}
\end{figure*}

In Figure~\ref{fig:ages_masses}, we compare our assembly times, and the stellar ages of \citet{gallazzi2005} to the median values of those found in the EAGLE simulation \citep{schaye2015}, and the SAM of H2015 as a function of stellar mass. In this figure, we record the median r-band weighted stellar age of a narrow stellar mass range ($\Delta\mathrm{\log{M_{*}/M_{\odot}}}=0.05$) of galaxies from the largest EAGLE simulation (Ref-L100N1504)  at $z=0.1$, and the millennium simulation (Henriques2015a..MRscPlanck1). We also trace the mass evolution of the most massive progenitors of theses galaxies to estimate an assembly redshift. 

Figure~\ref{fig:ages_masses} shows that the observations display a positive correlation between $t_{*}$, $t_{a}$ and stellar mass (as implied by the mass functions), assuming the relationships are of the form

\begin{equation}
t_{*} \propto \alpha\log{M_{*}},~~~t_{a} \propto \beta\log{M_{*}}
\label{eq:linear}
\end{equation}

where $\alpha$ and $\beta$ are the best fit linear slopes for $t_{*}$ and $t_{a}$ respectively. The H2015 model also reproduces the positive trend between stellar mass, $t_{*}$ and $t_{a}$, albeit with slightly flatter slopes. For observations, we find $\alpha_{obs}=4.25\pm0.55$, and $\beta_{obs}=2.46\pm0.30$ which are both steeper than the those implied for H2015 ($\alpha_{H2015} = 3.42\pm0.56$, $\beta_{H2015}=1.38\pm0.34$) although they agree to within $2\sigma$. This suggests the SAMs are doing a good job at reproducing the formation of stars and their assembly for the stellar mass ranges considered in this study, with a slight bias towards earlier formation.

The EAGLE simulation similarly reproduces the relationship between $t_{*}$ and $M_{*}$, although the value for $\alpha$ is even flatter than that of H2015 ($\alpha_{EAGLE}=2.19\pm0.39$). For the assembly time, EAGLE does not reproduce the trend at all, and instead has a $\beta$ consistent with 0 ($\beta_{EAGLE}=-0.17\pm0.40$). This is also seen in \citet{qu2017}, who performed a more robust analysis of the EAGLE simulation galaxy assembly, and who's median assembly times also indicate a flat relationship with $M_{*}$. 

At high masses ($M_{*}=10^{11.5}M_{\odot}$), compared to observations, massive EAGLE galaxies assemble their mass too quickly. There are also issues at the lowest mass where the discrepancy of stellar ages and assembly times in EAGLE at $\log{M_{*}/M_{\odot}}=10.0$ is significant (and also present in H2015, although not as discrepant in the SAM) and likely related to simulations over-producing low-mass galaxies at higher redshift \citep[see][]{weinmann2012, henriques2013, lacey2016}. 

\section{Discussion and Conclusions}

From Figure~\ref{fig:progenitors} and ~\ref{fig:ages}, we see a clear trend between $t_{a}$ and $t_{*}$ with stellar mass. More massive galaxies formed earlier, and at $\log{M_{*}/M_{\odot}}\geq10.5$, they also have stellar ages which are older than their respective assembly times, suggesting that a larger fraction of their stars formed ex-situ compared to lower mass galaxies. This picture implies that mergers are a more important component of stellar mass growth in massive galaxies, which is consistent with what is seen in previous studies \citep[e.g.,][]{naab2009, hopkins2010, vandokkum2010, trujillo2011, newman2012, hilz2013, mclure2013, vulcani2016, hill2017, mundy2017}.  In contrast, with $t_{a}\approx t_{*}$ for galaxies at $\log{M_{*}/M_{\odot}}<10.5$, almost all the stars can be attributed to in-situ formation.

Although the $t_{a}$ was not calculated explicitly, both \citet{patel2013a} and  \citet{vandokkum2013} used fixed cumulative number density arguments to calculate the stellar mass evolution as a function of redshift, from which a $z_{a}$ can be inferred. Using their fits for $M_{*}(z)$, for a $10^{11.2}\mathrm{M_{\odot}}$ galaxy, \citet{patel2013a} found an assembly redshift of $1.97$. Using our prescription, for the same galaxy mass, we would find $z_{a}=1.42$. \citet{vandokkum2013}, for a $10^{10.7}\mathrm{M_{\odot}}$ galaxy, find $z_{a}=1.35$ (and for which we would find $z_{a}=0.91$). Both studies find earlier assembly redshifts than we do. About half this redshift discrepancy is due different selection criteria (i.e. the use of a fixed cumulative number density instead of an evolving cumulative number density, where the former predicts higher mass progenitors (see \citealt{hill2017}), with the remainder due to the use of different mass functions. 

A comparison of our results to recent hydrodynamical simulations \citep{schaye2015} and semi-analytic models (H2015) show good agreement in the relationship of $t_{*}$ with stellar mass in all but the lowest mass bin (with the exception of the highest mass bin in the EAGLE simulation). This is especially impressive in EAGLE considering the models were not calibrated to reproduce stellar ages. The disagreement in $t_{*}$ in the lowest mass bin suggests either simulations are still forming stars too early, or, conversely the stellar ages in lower mass galaxies are underestimated. Using deep $(S/N(\AA)>50)$ spectroscopy of a handful of local group galaxies, \citet{sanchezblazquez2011} found that nearby barred-spiral galaxies were dominated by stars with ages on the order of $\sim10~Gyr$. Using deep, color-magnitude diagrams of local dwarfs, \citet{hidalgo2013} also found that the majority of stars in local dwarfs are between $9-10~\mathrm{Gyr}$ old. This is in apparent contradiction to the median ages found by \citet{gallazzi2005}. It is possible that the smaller local samples are not representative of the population as a whole. Conversely, the reverse could also be true and that the low-mass end of the galaxies from \citet{gallazzi2005} are also not representative. Alternatively, one way to resolve the discrepancy is to assume that there is a positive relation between stellar mass and age, which has a turnover at dwarf-galaxy stellar masses (although this seems unlikely). A more robust survey of low-mass, and hence low-surface brightness galaxies would be needed to address these issues.

If we assume that the mass-weighted stellar ages inferred from \citet{gallazzi2005} are correct, then the disagreement between observations and EAGLE of $t_{*}$ also folds into the assembly times, where we see more significant disagreement between EAGLE and our estimates (although with large scatter). EAGLE does not reproduce the positive correlation between $t_{a}$ and stellar mass, but instead predicts a flat relationship which might be related to the fact that EAGLE doesn't reproduce the GSMF \citep{furlong2015}. 

Considering the SAMs of a decade ago \citep[e.g.,][]{delucia2006}, there has been massive improvement, with the assembly times calculated from the most recent SAM (H2015) agreeing remarkably well with the observations (to within $2\sigma$), with a slight bias to early assembly times. Although there have been great improvements in recent modelling and simulation work in regards to reproducing the GSMF, these results suggests that there are potential systematic offsets which need to be addressed, and that the evolution of the baryonic component of low-mass galaxies has not been sufficiently decoupled from their host dark-matter halos. Observationally, there is an under-explored parameter space in regards to low-mass galaxies which are crucially needed to inform the simulations. 

\section{Summary}

In this Letter, we have measured the assembly time, and stellar ages from observations for four different mass descendant galaxies ($\log{M_{*}/M_{\odot}}=11.5,11.0,10.5,10.0$) at $z\sim0.1$ and find 

\begin{enumerate}
	\item The assembly times, and stellar ages decrease with decreasing stellar mass, consistent with cosmic 'down-sizing'. 
	\item The difference between $t_{a}$ and $t_{*}$ increases weakly with increasing stellar mass suggesting that massive galaxies form a larger fraction of their stars ex-situ compared to lower mass galaxies.
	\item $t_{a}$ and $t_{*}$ both increase with stellar mass, ranging from $\sim5-11~\mathrm{Gyr}$ in mass-weighted stellar age and $\sim5-9~\mathrm{Gyr}$ in assembly times. The SAM model of H2015 reproduces these trends to within $2\sigma$, albeit with slightly flatter relationships. EAGLE reproduces the positive correlation with $t_{*}$, but not with $t_{a}$ where EAGLE predicts no mass dependence on assembly times.
	\item The assembly times and stellar ages from the most recent SAM from the Millennium simulations \citep{henriques2015} are in good agreement with the observations, with a slight bias to earlier formation and assembly. 
\end{enumerate}

\section{Acknowledgements} 

We would like to thank Bruno Henriques, and James Trayford for deriving values from the H2015 and EAGLE catalogs, respectively. We are also grateful to Pieter van Dokkum and the anonymous referee whose comments greatly improved the presentation of this work. DM acknowledges the National Science Foundation under grant No. 1513473. This research has made use of NASA's Astrophysics Data System. 


\begin{thebibliography}{46}
\expandafter\ifx\csname natexlab\endcsname\relax\def\natexlab#1{#1}\fi

\bibitem[{{Behroozi} {et~al.}(2013){Behroozi}, {Marchesini}, {Wechsler},
  {Muzzin}, {Papovich}, \& {Stefanon}}]{behroozi2013}
{Behroozi}, P.~S., {Marchesini}, D., {Wechsler}, R.~H., {et~al.} 2013, \apjl,
  777, L10

\bibitem[{{Bernardi} {et~al.}(2017){Bernardi}, {Meert}, {Sheth}, {Fischer},
  {Huertas-Company}, {Maraston}, {Shankar}, \& {Vikram}}]{bernardi2017}
{Bernardi}, M., {Meert}, A., {Sheth}, R.~K., {et~al.} 2017, \mnras, 467, 2217

\bibitem[{{Bezanson} {et~al.}(2012){Bezanson}, {van Dokkum}, \&
  {Franx}}]{bezanson2012}
{Bezanson}, R., {van Dokkum}, P., \& {Franx}, M. 2012, \apj, 760, 62

\bibitem[{{Clauwens} {et~al.}(2016){Clauwens}, {Franx}, \&
  {Schaye}}]{clauwens2016}
{Clauwens}, B., {Franx}, M., \& {Schaye}, J. 2016, \mnras, 463, L1

\bibitem[{{De Lucia} {et~al.}(2006){De Lucia}, {Springel}, {White}, {Croton},
  \& {Kauffmann}}]{delucia2006}
{De Lucia}, G., {Springel}, V., {White}, S.~D.~M., {Croton}, D., \&
  {Kauffmann}, G. 2006, \mnras, 366, 499

\bibitem[{{Fontanot} {et~al.}(2007){Fontanot}, {Monaco}, {Silva}, \&
  {Grazian}}]{fontanot2007}
{Fontanot}, F., {Monaco}, P., {Silva}, L., \& {Grazian}, A. 2007, \mnras, 382,
  903

\bibitem[{{Fumagalli} {et~al.}(2016){Fumagalli}, {Franx}, {van Dokkum},
  {Whitaker}, {Skelton}, {Brammer}, {Nelson}, {Maseda}, {Momcheva}, {Kriek},
  {Labb{\'e}}, {Lundgren}, \& {Rix}}]{fumagalli2016}
{Fumagalli}, M., {Franx}, M., {van Dokkum}, P., {et~al.} 2016, \apj, 822, 1

\bibitem[{{Furlong} {et~al.}(2015){Furlong}, {Bower}, {Theuns}, {Schaye},
  {Crain}, {Schaller}, {Dalla Vecchia}, {Frenk}, {McCarthy}, {Helly},
  {Jenkins}, \& {Rosas-Guevara}}]{furlong2015}
{Furlong}, M., {Bower}, R.~G., {Theuns}, T., {et~al.} 2015, \mnras, 450, 4486

\bibitem[{{Gallazzi} {et~al.}(2005){Gallazzi}, {Charlot}, {Brinchmann},
  {White}, \& {Tremonti}}]{gallazzi2005}
{Gallazzi}, A., {Charlot}, S., {Brinchmann}, J., {White}, S.~D.~M., \&
  {Tremonti}, C.~A. 2005, \mnras, 362, 41

\bibitem[{{Goddard} {et~al.}(2017){Goddard}, {Thomas}, {Maraston}, {Westfall},  
	{Etherington}, {Riffel}, {Mallmann}, 
	{Zheng}, {Argudo-Fern{\'a}ndez},  {Lian}, 
	{Bershady}, {Bundy}, {Drory},  {Law}, 
	{Yan},  {Wake},  {Weijmans},  {Bizyaev}, 
	{Brownstein},  {Lane},  {Maiolino}, {Masters},
	{Merrifield}, {Nitschelm}, {Pan},  {Roman-Lopes},  
	{Storchi-Bergmann},  {Schneider}}]{goddard2017}
{Goddard}, D., {Thomas}, D., {Maraston}, C., {White}, {et~al.} 2017, \mnras, 466, 4731

\bibitem[{{Gonz{\'a}lez Delgado} {et~al.}(2017){Gonz{\'a}lez Delgado},
  {P{\'e}rez}, {Cid Fernandes}, {Garc{\'{\i}}a-Benito}, {L{\'o}pez
  Fern{\'a}ndez}, {Vale Asari}, {Cortijo-Ferrero}, {de Amorim}, {Lacerda},
  {S{\'a}nchez}, {Lehnert}, \& {Walcher}}]{gonzalezdelgado2017}
{Gonz{\'a}lez Delgado}, R.~M., {P{\'e}rez}, E., {Cid Fernandes}, R., {et~al.}
  2017, ArXiv e-prints

\bibitem[{{Grazian} {et~al.}(2015){Grazian}, {Fontana}, {Santini}, {Dunlop},
  {Ferguson}, {Castellano}, {Amorin}, {Ashby}, {Barro}, {Behroozi}, {Boutsia},
  {Caputi}, {Chary}, {Dekel}, {Dickinson}, {Faber}, {Fazio}, {Finkelstein},
  {Galametz}, {Giallongo}, {Giavalisco}, {Grogin}, {Guo}, {Kocevski},
  {Koekemoer}, {Koo}, {Lee}, {Lu}, {Merlin}, {Mobasher}, {Nonino}, {Papovich},
  {Paris}, {Pentericci}, {Reddy}, {Renzini}, {Salmon}, {Salvato}, {Sommariva},
  {Song}, \& {Vanzella}}]{grazian2015}
{Grazian}, A., {Fontana}, A., {Santini}, P., {et~al.} 2015, \aap, 575, A96

\bibitem[{{Henriques} {et~al.}(2015){Henriques}, {White}, {Thomas}, {Angulo},
  {Guo}, {Lemson}, {Springel}, \& {Overzier}}]{henriques2015}
{Henriques}, B.~M.~B., {White}, S.~D.~M., {Thomas}, P.~A., {et~al.} 2015,
  \mnras, 451, 2663

\bibitem[{{Henriques} {et~al.}(2013){Henriques}, {White}, {Thomas}, {Angulo},
  {Guo}, {Lemson}, \& {Springel}}]{henriques2013}
---. 2013, \mnras, 431, 3373

\bibitem[{{Hidalgo} {et~al.}(2013){Hidalgo}, {Monelli}, {Aparicio}, {Gallart},
  {Skillman}, {Cassisi}, {Bernard}, {Mayer}, {Stetson}, {Cole}, \&
  {Dolphin}}]{hidalgo2013}
{Hidalgo}, S.~L., {Monelli}, M., {Aparicio}, A., {et~al.} 2013, \apj, 778, 103

\bibitem[{{Hill} {et~al.}(2017){Hill}, {Muzzin}, {Franx}, {Clauwens},
  {Schreiber}, {Marchesini}, {Stefanon}, {Labbe}, {Brammer}, {Caputi}, {Fynbo},
  {Milvang-Jensen}, {Skelton}, {van Dokkum}, \& {Whitaker}}]{hill2017}
{Hill}, A.~R., {Muzzin}, A., {Franx}, M., {et~al.} 2017, \apj, 837, 147

\bibitem[{{Hilz} {et~al.}(2013){Hilz}, {Naab}, \& {Ostriker}}]{hilz2013}
{Hilz}, M., {Naab}, T., \& {Ostriker}, J.~P. 2013, \mnras, 429, 2924

\bibitem[{{Hopkins} {et~al.}(2010){Hopkins}, {Bundy}, {Hernquist}, {Wuyts}, \&
  {Cox}}]{hopkins2010}
{Hopkins}, P.~F., {Bundy}, K., {Hernquist}, L., {Wuyts}, S., \& {Cox}, T.~J.
  2010, \mnras, 401, 1099

\bibitem[{{Kauffmann} {et~al.}(2003){Kauffmann}, {Heckman}, {White}, {Charlot},
  {Tremonti}, {Brinchmann}, {Bruzual}, {Peng}, {Seibert}, {Bernardi},
  {Blanton}, {Brinkmann}, {Castander}, {Cs{\'a}bai}, {Fukugita}, {Ivezic},
  {Munn}, {Nichol}, {Padmanabhan}, {Thakar}, {Weinberg}, \&
  {York}}]{kauffmann2003}
{Kauffmann}, G., {Heckman}, T.~M., {White}, S.~D.~M., {et~al.} 2003, \mnras,
  341, 33

\bibitem[{{Lacey} {et~al.}(2016){Lacey}, {Baugh}, {Frenk}, {Benson}, {Bower},
  {Cole}, {Gonzalez-Perez}, {Helly}, {Lagos}, \& {Mitchell}}]{lacey2016}
{Lacey}, C.~G., {Baugh}, C.~M., {Frenk}, C.~S., {et~al.} 2016, \mnras, 462,
  3854

\bibitem[{{Marchesini} {et~al.}(2009){Marchesini}, {van Dokkum}, {F{\"o}rster
  Schreiber}, {Franx}, {Labb{\'e}}, \& {Wuyts}}]{marchesini2009}
{Marchesini}, D., {van Dokkum}, P.~G., {F{\"o}rster Schreiber}, N.~M., {et~al.}
  2009, \apj, 701, 1765

\bibitem[{{Marchesini} {et~al.}(2014){Marchesini}, {Muzzin}, {Stefanon},
  {Franx}, {Brammer}, {Marsan}, {Vulcani}, {Fynbo}, {Milvang-Jensen}, {Dunlop},
  \& {Buitrago}}]{marchesini2014}
{Marchesini}, D., {Muzzin}, A., {Stefanon}, M., {et~al.} 2014, \apj, 794, 65

\bibitem[{{McLure} {et~al.}(2013){McLure}, {Pearce}, {Dunlop}, {Cirasuolo},
  {Curtis-Lake}, {Bruce}, {Caputi}, {Almaini}, {Bonfield}, {Bradshaw},
  {Buitrago}, {Chuter}, {Foucaud}, {Hartley}, \& {Jarvis}}]{mclure2013}
{McLure}, R.~J., {Pearce}, H.~J., {Dunlop}, J.~S., {et~al.} 2013, \mnras, 428,
  1088

\bibitem[{{Mundy} {et~al.}(2017){Mundy}, {Conselice}, {Duncan}, {Almaini},
  {H{\"a}u{\ss}ler}, \& {Hartley}}]{mundy2017}
{Mundy}, C.~J., {Conselice}, C.~J., {Duncan}, K.~J., {et~al.} 2017, ArXiv
  e-prints

\bibitem[{{Muzzin} {et~al.}(2013){Muzzin}, {Marchesini}, {Stefanon}, {Franx},
  {McCracken}, {Milvang-Jensen}, {Dunlop}, {Fynbo}, {Brammer}, {Labb{\'e}}, \&
  {van Dokkum}}]{muzzin2013b}
{Muzzin}, A., {Marchesini}, D., {Stefanon}, M., {et~al.} 2013, \apj, 777, 18

\bibitem[{{Naab} {et~al.}(2009){Naab}, {Johansson}, \& {Ostriker}}]{naab2009}
{Naab}, T., {Johansson}, P.~H., \& {Ostriker}, J.~P. 2009, \apjl, 699, L178

\bibitem[{{Newman} {et~al.}(2012){Newman}, {Ellis}, {Bundy}, \&
  {Treu}}]{newman2012}
{Newman}, A.~B., {Ellis}, R.~S., {Bundy}, K., \& {Treu}, T. 2012, \apj, 746,
  162

\bibitem[{{Ownsworth} {et~al.}(2014){Ownsworth}, {Conselice}, {Mortlock},
  {Hartley}, {Almaini}, {Duncan}, \& {Mundy}}]{ownsworth2014}
{Ownsworth}, J.~R., {Conselice}, C.~J., {Mortlock}, A., {et~al.} 2014, \mnras,
  445, 2198

\bibitem[{{Papovich} {et~al.}(2015){Papovich}, {Labb{\'e}}, {Quadri}, {Tilvi},
  {Behroozi}, {Bell}, {Glazebrook}, {Spitler}, {Straatman}, {Tran}, {Cowley},
  {Dav{\'e}}, {Dekel}, {Dickinson}, {Ferguson}, {Finkelstein}, {Gawiser},
  {Inami}, {Faber}, {Kacprzak}, {Kawinwanichakij}, {Kocevski}, {Koekemoer},
  {Koo}, {Kurczynski}, {Lotz}, {Lu}, {Lucas}, {McIntosh}, {Mehrtens},
  {Mobasher}, {Monson}, {Morrison}, {Nanayakkara}, {Persson}, {Salmon},
  {Simons}, {Tomczak}, {van Dokkum}, {Weiner}, \& {Willner}}]{papovich2015}
{Papovich}, C., {Labb{\'e}}, I., {Quadri}, R., {et~al.} 2015, \apj, 803, 26

\bibitem[{{Patel} {et~al.}(2013{\natexlab{a}}){Patel}, {van Dokkum}, {Franx},
  {Quadri}, {Muzzin}, {Marchesini}, {Williams}, {Holden}, \&
  {Stefanon}}]{patel2013a}
{Patel}, S.~G., {van Dokkum}, P.~G., {Franx}, M., {et~al.} 2013{\natexlab{a}},
  \apj, 766, 15

\bibitem[{{Patel} {et~al.}(2013{\natexlab{b}}){Patel}, {Fumagalli}, {Franx},
  {van Dokkum}, {van der Wel}, {Leja}, {Labb{\'e}}, {Brammer}, {Skelton},
  {Momcheva}, {Whitaker}, {Lundgren}, {Muzzin}, {Quadri}, {Nelson}, {Wake}, \&
  {Rix}}]{patel2013b}
{Patel}, S.~G., {Fumagalli}, M., {Franx}, M., {et~al.} 2013{\natexlab{b}},
  \apj, 778, 115

\bibitem[{{P{\'e}rez-Gonz{\'a}lez} {et~al.}(2008){P{\'e}rez-Gonz{\'a}lez},
  {Rieke}, {Villar}, {Barro}, {Blaylock}, {Egami}, {Gallego}, {Gil de Paz},
  {Pascual}, {Zamorano}, \& {Donley}}]{perezgonzalez2008}
{P{\'e}rez-Gonz{\'a}lez}, P.~G., {Rieke}, G.~H., {Villar}, V., {et~al.} 2008,
  \apj, 675, 234

\bibitem[{{Qu} {et~al.}(2017){Qu}, {Helly}, {Bower}, {Theuns}, {Crain},
  {Frenk}, {Furlong}, {McAlpine}, {Schaller}, {Schaye}, \& {White}}]{qu2017}
{Qu}, Y., {Helly}, J.~C., {Bower}, R.~G., {et~al.} 2017, \mnras, 464, 1659

\bibitem[{{Rodriguez-Gomez} {et~al.}(2016){Rodriguez-Gomez}, {Pillepich},
  {Sales}, {Genel}, {Vogelsberger}, {Zhu}, {Wellons}, {Nelson}, {Torrey},
  {Springel}, {Ma}, \& {Hernquist}}]{rodriguezgomez2016}
{Rodriguez-Gomez}, V., {Pillepich}, A., {Sales}, L.~V., {et~al.} 2016, \mnras,
  458, 2371

\bibitem[{{Rodr{\'{\i}}guez-Puebla} {et~al.}(2017){Rodr{\'{\i}}guez-Puebla},
  {Primack}, {Avila-Reese}, \& {Faber}}]{rodriguezpuebla2017}
{Rodr{\'{\i}}guez-Puebla}, A., {Primack}, J.~R., {Avila-Reese}, V., \& {Faber},
  S.~M. 2017, \mnras, 470, 651

\bibitem[{{S{\'a}nchez-Bl{\'a}zquez} {et~al.}(2011){S{\'a}nchez-Bl{\'a}zquez},
  {Ocvirk}, {Gibson}, {P{\'e}rez}, \& {Peletier}}]{sanchezblazquez2011}
{S{\'a}nchez-Bl{\'a}zquez}, P., {Ocvirk}, P., {Gibson}, B.~K., {P{\'e}rez}, I.,
  \& {Peletier}, R.~F. 2011, \mnras, 415, 709

\bibitem[{{Schaye} {et~al.}(2015){Schaye}, {Crain}, {Bower}, {Furlong},
  {Schaller}, {Theuns}, {Dalla Vecchia}, {Frenk}, {McCarthy}, {Helly},
  {Jenkins}, {Rosas-Guevara}, {White}, {Baes}, {Booth}, {Camps}, {Navarro},
  {Qu}, {Rahmati}, {Sawala}, {Thomas}, \& {Trayford}}]{schaye2015}
{Schaye}, J., {Crain}, R.~A., {Bower}, R.~G., {et~al.} 2015, \mnras, 446, 521

\bibitem[{{Sparre} {et~al.}(2015){Sparre}, {Hayward}, {Springel},
  {Vogelsberger}, {Genel}, {Torrey}, {Nelson}, {Sijacki}, \&
  {Hernquist}}]{sparre2015}
{Sparre}, M., {Hayward}, C.~C., {Springel}, V., {et~al.} 2015, \mnras, 447,
  3548

\bibitem[{{Thomas} {et~al.}(2010){Thomas}, {Maraston}, {Schawinski}, {Sarzi},
  \& {Silk}}]{thomas2010}
{Thomas}, D., {Maraston}, C., {Schawinski}, K., {Sarzi}, M., \& {Silk}, J.
  2010, \mnras, 404, 1775

\bibitem[{{Torrey} {et~al.}(2015){Torrey}, {Wellons}, {Machado}, {Griffen},
  {Nelson}, {Rodriguez-Gomez}, {McKinnon}, {Pillepich}, {Ma}, {Vogelsberger},
  {Springel}, \& {Hernquist}}]{torrey2015}
{Torrey}, P., {Wellons}, S., {Machado}, F., {et~al.} 2015, \mnras, 454, 2770

\bibitem[{{Trujillo} {et~al.}(2011){Trujillo}, {Ferreras}, \& {de La
  Rosa}}]{trujillo2011}
{Trujillo}, I., {Ferreras}, I., \& {de La Rosa}, I.~G. 2011, \mnras, 415, 3903

\bibitem[{{van Dokkum} {et~al.}(2010){van Dokkum}, {Whitaker}, {Brammer},
  {Franx}, {Kriek}, {Labb{\'e}}, {Marchesini}, {Quadri}, {Bezanson},
  {Illingworth}, {Muzzin}, {Rudnick}, {Tal}, \& {Wake}}]{vandokkum2010}
{van Dokkum}, P.~G., {Whitaker}, K.~E., {Brammer}, G., {et~al.} 2010, \apj,
  709, 1018

\bibitem[{{van Dokkum} {et~al.}(2013){van Dokkum}, {Leja}, {Nelson}, {Patel},
  {Skelton}, {Momcheva}, {Brammer}, {Whitaker}, {Lundgren}, {Fumagalli},
  {Conroy}, {F{\"o}rster Schreiber}, {Franx}, {Kriek}, {Labb{\'e}},
  {Marchesini}, {Rix}, {van der Wel}, \& {Wuyts}}]{vandokkum2013}
{van Dokkum}, P.~G., {Leja}, J., {Nelson}, E.~J., {et~al.} 2013, \apjl, 771,
  L35

\bibitem[{{van Dokkum} {et~al.}(2014){van Dokkum}, {Bezanson}, {van der Wel},
  {Nelson}, {Momcheva}, {Skelton}, {Whitaker}, {Brammer}, {Conroy},
  {F{\"o}rster Schreiber}, {Fumagalli}, {Kriek}, {Labb{\'e}}, {Leja},
  {Marchesini}, {Muzzin}, {Oesch}, \& {Wuyts}}]{vandokkum2014}
{van Dokkum}, P.~G., {Bezanson}, R., {van der Wel}, A., {et~al.} 2014, \apj,
  791, 45

\bibitem[{{Vulcani} {et~al.}(2016){Vulcani}, {Marchesini}, {De Lucia},
  {Muzzin}, {Stefanon}, {Brammer}, {Labb{\'e}}, {Le F{\`e}vre}, \&
  {Milvang-Jensen}}]{vulcani2016}
{Vulcani}, B., {Marchesini}, D., {De Lucia}, G., {et~al.} 2016, \apj, 816, 86

\bibitem[{{Weinmann} {et~al.}(2012){Weinmann}, {Pasquali}, {Oppenheimer},
  {Finlator}, {Mendel}, {Crain}, \& {Macci{\`o}}}]{weinmann2012}
{Weinmann}, S.~M., {Pasquali}, A., {Oppenheimer}, B.~D., {et~al.} 2012, \mnras,
  426, 2797

\bibitem[{{Wellons} \& {Torrey}(2016)}]{wellons2016b}
{Wellons}, S., \& {Torrey}, P. 2016, ArXiv e-prints

\end{thebibliography}

\end{document}